\documentclass{llncs}

\usepackage{makeidx}
\usepackage{comment}
\usepackage{amsmath,amssymb,graphicx}
\usepackage{theorem}
\usepackage{here}
\usepackage{url}
\usepackage{algorithm,algorithmicx,algpseudocode}

\usepackage[version=3]{mhchem} 
\usepackage[T1]{fontenc}       

\newtheorem{thrm}{Theorem}

\title{Scalable Similarity Search for\\ Molecular Descriptors\thanks{This work was supported the JST PRESTO program and the Academy of Finland through grants 294143 and 2845984.}}

\author{Yasuo Tabei\inst{1}
        \and Simon J. Puglisi\inst{2}}
\institute{RIKEN Center for Advanced Intelligence Project\\
\email{yasuo.tabei@riken.jp}
\and
Helsinki Institute for Information Technology,\\
Department of Computer Science,\\
University of Helsinki\\
\email{puglisi@cs.helsinki.fi}
}

\begin{document}
\maketitle

\begin{abstract}
Similarity search over chemical compound databases is a fundamental task in the discovery and design of novel drug-like molecules. Such databases often encode molecules as non-negative integer vectors, called {\em molecular descriptors}, which represent rich information on various molecular properties. 
While there exist efficient indexing structures for searching databases of {\em binary} vectors, solutions for more general integer vectors are in their infancy. 
In this paper we present a time- and space-efficient index for the problem that we call the {\em succinct intervals-splitting tree algorithm for molecular descriptors (SITAd)}. Our approach extends efficient methods for binary-vector databases, and uses ideas from succinct data structures.
Our experiments, on a large database of over $40$ million compounds, show SITAd significantly outperforms alternative approaches in practice.
\end{abstract}

\section{Introduction}
Molecules that are chemically similar tend to have a similar molecular function. 
The first step in predicting the function of a new molecule is, therefore, to conduct 
a similarity search for the molecule in huge databases of molecules with known properties and functions. 
Current molecular databases store vast numbers of chemical compounds. 
For example, the PubChem database in the National Center for Biotechnology Information (NCBI) files more than $40$ million molecules. 
Because the size of the whole chemical space\cite{Kei07} is said to be approximately $10^{60}$, 
molecular databases are growing and are expected to grow substantially in the future.
There is therefore a strong need to develop scalable methods for rapidly searching for molecules
that are similar to a previously unseen target molecule.

A {\em molecular fingerprint}, defined as a binary vector, is a standard representation of molecules in chemoinformatics\cite{Todeschini2002}. 
In practice the fingerprint representation of molecules is in widespread use~\cite{Chen09,Chen05} because it conveniently encodes the presence or absence of molecular substructures and functions. 
Jaccard similarity, also called Tanimoto similarity, is the {\em de facto} standard measure~\cite{Leach2007} to evaluate similarities between compounds represented as fingerprints in chemoinformatics and pharmacology.
To date, a considerable number of similarity search methods for molecular fingerprints using Jaccard similarity have been proposed~\cite{Tabei12,Tabei13,Thomas09,Nasr12}.
Among them, the {\em succinct intervals-splitting tree} (SITA)~\cite{Tabei12} is the fastest method that is also capable of dealing with large databases.
Despite the current popularity of the molecular fingerprint representation in cheminformatics, because it is only
a binary feature vector, it has a severely limited ability to distinguish between molecules, and so similarity search is often ineffective~\cite{Kristensen11}. 

A {\em molecular descriptor}, defined as a non-negative integer vector, is a powerful representation of molecules and enables storing richer information on various properties of molecules than a fingerprint does. 
Representative descriptors are LINGO~\cite{Vidal05} and KCF-S~\cite{Kotera12}. 
Recent studies have shown descriptor representations of molecules to be significantly better than fingerprint representations for predicting and interpreting molecular functions~\cite{Kotera13} and interactions~\cite{Sawada14}.
Although similarity search using descriptor representations of molecules is expected to become common in the near future, 
no efficient method for a similarity search with descriptors has been proposed so far. 
Kristensen et al.~\cite{Kristensen11} presented a fast similarity search method for molecular descriptors using an inverted index. The inverted index however consumes a large amount of memory when applied to large molecular databases. Of course one can compress the inverted index to reduce memory usage, but then the overhead of decompression at query time results in slower performance.
An important open challenge is thus to develop similarity search methods for molecular descriptors that are simultaneously fast and have a small memory footprint.

We present a novel method called SITAd by modifying the idea behind SITA. 
SITAd efficiently performs similarity search of molecular descriptors using generalized Jaccard similarity.
By splitting a database into clusters of descriptors using upperbound information of generalized Jaccard similarity 
and then building binary trees that recursively split descriptors on each cluster, 
SITAd can effectively prune out useless portions of the search space.
While providing search times as fast as inverted index-based approaches, SITAd requires substantially less memory by exploiting tools from succinct data structures, in particular rank dictionaries~\cite{Kark14a} and wavelet trees~\cite{Gro03}. 
SITAd efficiently solves range maximum queries (RMQ) many times in similarity searches by using fast RMQ data structures~\cite{Bender05} that are necessary
for fast and space-efficient similarity searches. 
By synthesizing these techniques, SITAd's time complexity is {\em output-sensitive}. 
That is, the greater the desired similarity with the query molecule is, the faster SITAd returns answers. 

To evaluate SITAd, we performed retrieval experiments over a huge database of more than $40$ million chemical compounds from the PubChem database. 
Our results demonstrate SITAd to be significantly faster and more space efficient than state-of-the-art methods.

\section{Similarity search problem for molecular descriptors}
We now formulate a similarity search problem for molecular descriptors. 
A molecular descriptor is a fixed-dimension vector, each dimension of which is a non-negative integer. 
It is conceptually equivalent to the set 
that consists of pairs $(d:f)$ of index $d$ and weight $f$ such that the $d$-th dimension of the descriptor is a non-zero value $f$. 
Let $D$ be a dimension with respect to the vector representation of descriptors. 
For clarity, notations $x_i$ and $q$ denote $D$ dimension vector representation of molecular descriptors, while $W_i$ and $Q$ correspond to their set representation. 
$|W_i|$ denotes the cardinality of $W_i$, i.e., the number of elements in $W_i$. 
The Jaccard similarity for two vectors $x$ and $x^\prime$ is defined as $J(x,x^\prime)=\frac{x\cdot x^\prime}{||x||_2^2 + ||x^\prime||_2^2 - x\cdot x^\prime}$ where $||x||_2$ is the $L_2$ norm. 
For notational convenience, we let $J(W,W^\prime)$ represent $J(x,x^\prime)$ of $x$ and $x^\prime$ that correspond respectively to sets $W$ and $W^\prime$. 
Given a query compound $Q$, the similarity search task is to retrieve from the database of $N$ compounds all the identifiers $i$ of descriptors $W_i$ whose Jaccard similarity between $W_i$ and $Q$ is no less than $\epsilon$, i.e., the set $I_N = \{i\in \{1,2,...,N\}; J(W_i,Q) \geq \epsilon\}$.

\section{Method}
Our method splits a database into blocks of descriptors with the same squared norm and 
searches descriptors similar to a query in a limited number of blocks satisfying a similarity constraint. 
Our similarity constraint depends on Jaccard similarity threshold $\epsilon$.
The larger $\epsilon$ is, the smaller the number of selected blocks is. 
A standard method is to compute the Jaccard similarity between the query and each descriptor in the selected blocks, and then check whether or not the similarity is larger than $\epsilon$. 
However, such pairwise computation of Jaccard similarity is prohibitively time consuming.  
Our method builds an intervals-splitting tree for each block of descriptors and searches descriptors similar to a query by pruning useless portions of the search space in the tree.

\subsection{Database partitioning}
We relax the solution set $I_N$ for fast search using the following theorem. 
\begin{thrm}
  If $J(x,q) \geq \epsilon$, then $\epsilon ||q||_2^2 \leq ||x||_2^2 \leq ||q||_2^2/\epsilon$. 
\end{thrm}
{\bf Proof}~~~
$J(x,q)\geq\epsilon$ is equivalent to $|x\cdot q| \geq \frac{\epsilon}{1+\epsilon}(||x||_2^2+||q||_2^2)$. 
By the Cauchy-Schwarz inequality $||x||_2||q||_2 \geq |x\cdot q|$, we obtain $||x||_2||q||_2 \geq \frac{\epsilon}{1+\epsilon}(||x||_2^2+||q||_2^2)$.
When $||x||_2 \geq ||q||_2$, we get $||x||_2^2 \geq \epsilon ||q||_2^2$. 
Otherwise, we get $||q||^2_2/\epsilon \geq ||x||_2^2$.
Putting these results together, the theorem is obtained.

The theorem indicates that $I_1=\{i\in \{1,2,...,N\}; \epsilon||q||_2^2 \leq ||x_i||_2^2 \leq ||q||_2^2/\epsilon\}$ must contain all elements in $I_N$, i.e., $I_N \subseteq I_1$. 
This means a descriptor identifier (ID) that is not in $I_1$ is never a member of $I_N$. 
Such useless descriptors can be efficiently excluded by partitioning the database into blocks, each of which contains descriptor IDs 
with the same squared norm. 
More specifically, let block $B^c=\{i\in \{1,2,...,N\};||x_i||^2_2=c\}$ be the block containing all the descriptors in the database with squared norm $c$.
Searching descriptors for a query needs to examine no element in $B^c$ if either $c<\epsilon ||q||_2^2$ or $c>||q||_2^2/\epsilon$ holds. 

\subsection{Intervals-splitting tree for efficient similarity search}
Once blocks $B^c$ satisfying $\epsilon ||q||^2_2 \leq c \leq ||q||^2_2/\epsilon$ are selected, 
SITAd is able to bypass one-on-one computations of Jaccard similarity between each descriptor in $B^c$ and a query $q$. 

A binary tree $T^c$ called an intervals-splitting tree is built on each $B^c$ beforehand. 
When a query $q$ is given,  $T^c$ is traversed with a pruning scheme to efficiently select all the descriptor IDs with squared norm $c$ whose Jaccard similarity to query $q$ is no less than $\epsilon$. 
Each node in $T^c$ represents a set of descriptor IDs by using an interval of $B^c$. 
Let $B^c[i]$ be the $i$-th descriptor ID in $B^c$ and $I^c_v$ be the interval of node $v$. 
Node $v$ with interval $I^c_v=[s,e]$ contains descriptor IDs $B^c[s],B^c[s+1],\cdots,B^c[e]$. 
The interval of a leaf is of the form $[s,s]$, indicating that the leaf has only one ID. 
The interval of the root is $[1,|B^c|]$.

\begin{figure}[t]
\begin{center}
\begin{tabular}{c}
\includegraphics[width=0.95\textwidth]{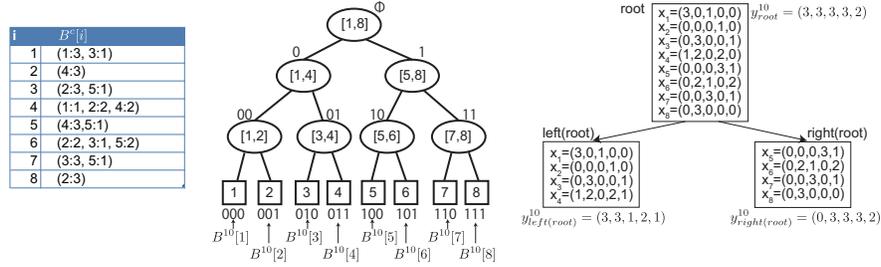}
\end{tabular}
\end{center}
\vspace{-0.5cm}
\caption{Descriptors in block $B^{10}$ (left), intervals-splitting tree $T^c$ (middle) and $T^c$'s first two levels (right). The root interval $[1,8]$ is split into $[1,4]$ and $[5,8]$ for the left and right children. Each node $v$ has a summary descriptor $y_v$ for the descriptors in its interval.}
\label{fig:bintree}
\end{figure}

Let $\mbox{{\em left}}(v)$ and $\mbox{{\em right}}(v)$ be the left and right children of node $v$ with interval $I^c_v=[s,e]$, respectively. 
When these children are generated, $I^c_v=[s,e]$ is partitioned into disjoint segments $I^c_{\mbox{{\scriptsize{\em left}}}(v)}=[s,\lfloor(s+e)/2\rfloor]$ and $I^c_{\mbox{{\scriptsize{\em right}}}(v)}=[\lfloor(s+e)/2\rfloor +1, e]$.
The procedure of splitting the interval is recursively applied from the root to the leaves (see the middle and right of Figure~\ref{fig:bintree} illustrating intervals and 
sets of descriptors at the root and its children).

Each node $v$ is identified by a bit string (e.g., $v=010$) 
indicating the path from the root to $v$; ``0'' and ``1'' denote the selection of left and right children, respectively. 
At each leaf $v$, the index of $B^c$ is calculated by $\mbox{{\em int}}(v)+1$, where $\mbox{{\em int}}(\cdot)$ converts a bit string to its corresponding integer (see the middle of Figure~\ref{fig:bintree}). 

\subsection{Pruning the search space using summary descriptors}
\label{sec:search}
Given query $q$, SITAd recursively traverses $T^c$ from the root in a depth-first manner. 
If SITAd reaches a leaf and its descriptor is similar to $q$, the ID of that descriptor is included as one solution. 
To avoid traversing the whole $T^c$, we present a scheme to prune subtrees of nodes if all the descriptors for the nodes are deemed not to be sufficiently similar to query $q$. 

The pruning is performed on node $v$ by using $D$ dimension descriptor $y_v$, which summarizes the information on descriptors in $I_v$, 
and is used for computing the upperbound of the Jaccard similarity between query $q$ and $X_{B^c[i]}$ for any $i \in I_v$.
The $d$-th dimension $y_v[d]$ of $y_v$ is defined as the maximum value among $x_{B^c[i]}[d]$ for any $i\in I_v$, i.e., $y_v[d]=\max_{i\in I_v}{x_{B^c[i]}[d]}$. Thus
$y_v=(\max_{i \in I_v}x_{B^c[i]}[1]$, $\max_{i\in I_v}x_{B^c[i]}[2], ..., \max_{i\in I_v}x_{B^c[i]}[D])$.
When $T^c$ is built, $y_v$ is computed.
(see the right of Figure~\ref{fig:bintree}, which represents $y_v$ in the first two-level nodes of $T^{10}$).

Assume that SITAd checks descriptors in $B^c$ and traverses $T^c$ in the depth-first manner. 
$||x_{B^c[i]}||^2=c$ holds in any descriptor in $B^c$. 
The following equivalent constraint is derived from Jaccard similarity: 
\begin{eqnarray}
 &J(x_{B^c[i]},q)&=\frac{x_{B^c[i]} \cdot q}{||x_{B^c[i]}||^2_2 + ||q||^2_2 - x_{B^c[i]}\cdot q} \geq \epsilon  \nonumber \\
\Longleftrightarrow & x_{B^c[i]} \cdot q & \geq \frac{\epsilon}{1+\epsilon}(||x_{B^c[i]}||^2_2 + ||q||^2_2) = \frac{\epsilon}{1+\epsilon}(c + ||q||^2_2) \nonumber
\end{eqnarray}
Since $y^c_v \cdot q \geq x_{B^c[i]}\cdot q$ holds for any $i \in I^c_v$, 
SITAd examines the constraint at each node $v$ in $T^c$ and checks whether or not the following condition,    
\begin{eqnarray}
  \sum_{(d:f) \in Q} y_v^c[d]f \geq \frac{\epsilon}{1+\epsilon}(c+||q||^2),
\label{eq:node}
\end{eqnarray} 
holds at each node $v$.
If the inequality does not hold at node $v$, SITAd safely prunes the subtrees rooted at $v$, because there are no descriptors similar to $q$ in leaves under $v$. As we shall see, this greatly improves 
search efficiency in practice. 
Algorithm~\ref{alg:1} shows the pseudo-code of SITAd.  

\begin{algorithm}[t]
\caption{{\small Algorithm for finding similar descriptors to query $q$.}}
\label{alg:1}
\begin{algorithmic}[1]
{\small
\Function{Search}{$q$}
\For{$c$ satisfying $\epsilon ||q||^2_2 \leq c \leq ||q||^2_2/\epsilon$}
\State $k \leftarrow \frac{\epsilon}{1+\epsilon}(c + ||q||^2_2)$, $I^c_{root} \leftarrow [1, |B^c|]$, $v \leftarrow \phi$
\State Recursion($v$,$I^c_{root}$,$q$,$c$)
\EndFor
\EndFunction
\Function{Recursion}{$v$,$I^c_v$,$q$,$c$}
\If{$\sum_{(d:f)\in Q}y^c_v[d]f < k$} \Comment{$Q$ : set representation of $q$}
\State \Return
\EndIf
\If{$|v|=\lceil \log|B_c|\rceil$} \Comment{Leaf Node}
\State{Output index $B^c[int(v)+1]$}
\EndIf
\State Recursive($v+'0'$,$[s,\lfloor (s+e)/2 \rfloor]$,$q$,$c$) \Comment{To left child}
\State Recursive($v+'1'$,$[\lfloor (s+e)/2\rfloor+1,e]$,$q$,$c$) \Comment{To right child}
\EndFunction
}
\end{algorithmic}
\end{algorithm}

\subsection{Search time and memory}
SITAd efficiently traverses $T^c$ by pruning its useless subtrees. 
Let $\tau$ be the numbers of traversed nodes. 
The search time for query $Q$ is $O(\tau|Q|)$. 
In particular, SITAd is efficient for larger $\epsilon$, because 
more nodes in $T^c$ are pruned. 

A crucial drawback of SITAd is that $T^c$ requires 
$O(D\log{M}|B^c|\log{(|B^c|)})$ space for each $c$, the dimension $D$ of descriptors and the maximum value $M$ among all weight values in descriptors. 
Since $D$ is large in practice, SITAd consumes a large amount of memory. 
The next two subsections describe approaches to reduce the memory usage 
while retaining query-time efficiency.

\subsection{Space reduction using inverted index} \label{sec:inv}
To reduce the large amount of space needed to store summary descriptors, 
we use an inverted index that enables computing an upperbound on descriptor similarity.
The inverted index itself does not always reduce the memory requirement. 
However, SITAd compactly maintains the information in a {\em rank dictionary}, significantly decreasing memory usage. 

\begin{figure}[t]
\begin{center}
\begin{tabular}{c}
\includegraphics[width=0.9\textwidth]{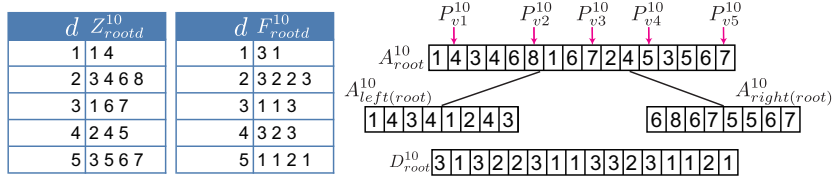}
\end{tabular}
\end{center}
\vspace{-0.5cm}
\caption{Example of $Z^{c}_{vd}$, $F^{c}_{vd}$, $A^c_{v}$ and $D^c_{v}$ for $T^c$.}
\label{fig:invtree}
\end{figure}

We use two kinds of inverted indexes for separately storing index and weight pairs in descriptors. 
One is an associative array that maps each index $d$ to the set of all descriptor IDs that contain pairs $(d:f)$ 
of index $d$ and any weight $f(\neq 0)$ at each node $v$. 
Let $Z^c_{vd}=\{i\in I^c_v; (d:f)\in W_{B^c[i]}~\mbox{for any}~ f(\neq 0)\}$ for index $d$, (i.e., all IDs of a descriptor containing $d$ with any weight $f$ 
in any pair $(d:f)$ within $I^c_v$.
The inverted index for storing indexes at node $v$ in $T^c$ is a one-dimensional array that concatenates all $Z^c_{vd}$ in 
ascending order of $d$ and is defined as $A^c_{v}=Z^c_{v1}\cup Z^c_{v2}\cup \cdots \cup Z^c_{vD}$. 
Figure~\ref{fig:invtree} shows $Z^{10}_{rootd}$ and the first two levels of the inverted indexes $A^{10}_{root}$, $A^{10}_{left(root)}$ and $A^{10}_{right(root)}$ in $T^{10}$ in Figure~\ref{fig:bintree}.

The other kind of inverted index is also an associative array that maps each index $d$ to the set of all weights that are paired with $d$. Let $F^c_{vd}=\{f; (d,f)\in W_{B^c[i]}, i\in I^c_v \}$ for index $d$ (i.e., all weights that are paired with $d$ within $I^c_v$). 
The inverted index for storing weights at node $v$ in $T^c$ is a one-dimensional array that concatenates all $F^c_{vd}$ in ascending order of $d$ 
and is defined as $E^c_{v}=F^c_{v1}\cup F^c_{v2}\cup \cdots \cup F^c_{vD}$. 
We build $E^c_{v}$ at only the root, i.e., $E^c_{root}=F^c_{root1}\cup F^c_{root2}\cup \cdots \cup F^c_{rootD}$. 
Figure~\ref{fig:invtree} shows an example of $E^{10}_{root}$ in $T^{10}$ in Figure~\ref{fig:bintree}. 

Let $P^c_{vd}$ indicate the ending position of $Z^c_{vd}$ and $F^c_{vd}$ on $A^c_v$ and $E^c_v$ for each $d\in [1,D]$, i.e., 
$P^c_{v0}=0$ and $P^c_{vd}=P^c_{v(d-1)}+|Z^c_{vd}|$ for $d=1,2,...,D$. 
If all descriptors at node $v$ do not have any pair $(d:f)$ of index $d$ and any weight $f(\neq 0)$, then $P^c_{vd}=P^c_{v(d+1)}$ holds. 

When searching for descriptors similar to query $Q=(d_1:f_1,d_2:f_2,...,d_m:f_m)$ in $T^c$, 
we traverse $T^c$ from the root. 
At each node, we set $s_{vj}=P^c_{v(d_{(j-1)})}+1$ and $t_{vj}=P^c_{v{d_j}}$ for $j=1,2,...,m$. 
If $s_{vj} \leq t_{vj}$ holds, there is at least one descriptor that contains $d_j$ because of $A^c_v$'s property. 
Otherwise, no descriptor at $v$ contains $d_j$. 
We check the following constraint, which is equivalent to condition~(\ref{eq:node}) as $\sum_{(d:f)\in Q}y^c_v[d]f \geq \frac{\epsilon}{(1+\epsilon)}(c+||q||^2)$ at each node $v$, 
\begin{eqnarray} \label{eq:cost}
\sum_{j=1}^m I[s_{vj} \leq t_{vj}]\cdot \max E_{root}^c[s_{vj},t_{vj}]\cdot f_j \geq \frac{\epsilon}{1+\epsilon}(c+||q||_2^2), 
\end{eqnarray} 
where $I[\mbox{{\em cond}}]$ is the indicator function that returns one if $cond$ is true and zero otherwise and  
$\max E_{root}^c[s_{vj},t_{vj}]$ returns the maximum value in subarray $E_{root}^c[s_{vj},t_{vj}]$.
For example in Figure~\ref{fig:invtree}, for $Q=(1:3,3:1,4:2)$ and $A^{10}_{root}$, 
$I[1 \leq 2]\cdot \max \{3,1\} \cdot 3 + I[7\leq 9] \cdot  \max \{1,1,3\} \cdot 1 + I[10 \leq 12] \cdot \max \{3,2,3\} \cdot 2 = 18$

A crucial observation is that computing constraint~(\ref{eq:cost}) needs $s_{vj}$, $t_{vj}$ and $\max D_{root}^c[s_{vj},t_{vj}]$ at each node $v$.
If we compute $s_{vj}$ and $t_{vj}$ at each node $v$, we can omit $A^c_{v}$, resulting in a huge memory reduction. 
We compute $s_{vj}$ and $t_{vj}$ using rank dictionaries.
The problem of computing $\max D_{root}^c[s_{vj},t_{vj}]$ is called a {\em range maximum query (RMQ)}.
Rank dictionaries and RMQ data structures are reviewed in the next section.

\subsection{Rank dictionaries and RMQ data structures}

\subsubsection{Rank dictionary}
A rank dictionary\cite{Ram02} is a data structure built over a bit array $B$ of length $n$. It supports the rank query $rank_c(B,i)$, which returns the number of occurrences of $c\in \{0,1\}$ in $B[1,i]$. 
Although naive approaches require the $O(n)$ time to compute a rank, several data structures with only $n+o(n)$ bits storage have been presented to achieve $O(1)$ query time~\cite{Okanohara07}.
We employ {\em hybrid bit vectors}~\cite{Kark14a} (which are compressed rank dictionaries) to calculate $I[s_{vj} \leq t_{vj}]$ in eq~(\ref{eq:cost}) with $O(1)$ time and only $n+o(n)$ bits (and sometimes much less). 

\subsubsection{RMQ data structures}
The RMQ problem for an array $D$ of length $n$ is defined as follows:
for indices $i$ and $j$ between $1$ and $n$, query $RMQ_{E}[i,j]$ returns the index of the largest element in subarray $E[i,j]$.
An RMQ data structure is built by preprocessing $E$ and is used for efficiently solving the RMQ problem. 

A naive data structure is simply a table storing $RMQ_{E}(i,j)$ for all possible pairs $(i,j)$ such that $1\leq i<j\leq n$. This takes $O(n^2)$ preprocessing time and space, and
it solves the RMQ problem in $O(1)$ query time.
An $O(n)$ preprocessing time and $O(1)$ query time data structure has been proposed\cite{Bender05} that uses $\frac{n\log{n}}{2}+n\log{M}+2n$ bits of space. 
RMQ data structure $U^c$ for each $c\in [1,D]$ is built for $E^c_{root}$ in $O(N^c)$ preprocessing time where $N^c$ is the total number of pairs $(d:f)$ in $B^c$, 
i.e., $N^c=\sum_{i\in B^c}|W_i|$. Then, $\max{E^c_{root}[s_{vj},t_{vj}]}$ in eq~(\ref{eq:cost}) can be computed using $U^c$ in $O(1)$ time. 

\subsection{Similarity search using rank dictionaries and RMQs} 
\begin{figure}[t]
\begin{center}
\begin{tabular}{c}
\includegraphics[width=0.8\textwidth]{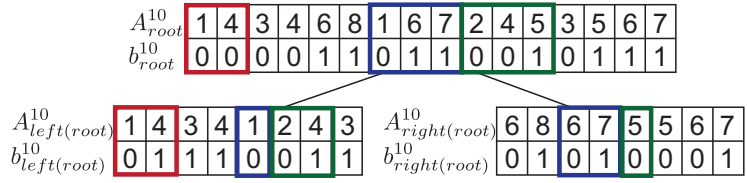}
\end{tabular}
\end{center}
\vspace{-0.5cm}
\caption{First two levels of wavelet tree in Figure~\ref{fig:invtree}.}
\label{fig:sitad}
\end{figure}

At the heart of SITAd is the wavelet tree, a succinct data structure usually applied to string data~\cite{Gro03}.
SITAd stores only rank dictionaries and an RMQ data structure without maintaining $A_v^c$ and $E_v^c$ in memory.
Thus, SITAd can compute constraint~(\ref{eq:cost}) in a space-efficient manner. 

A wavelet tree is a collection of rank dictionaries to update interval at each node. 
Let $b_v^c$ be a bit array of length $|A^c_v|$ at node $v$.
Let $I^c_v=[a,b]$ be an interval of node $v$, and let $\mbox{{\em left}}(v)$ (resp. $\mbox{{\em right}}(v)$) be the left child (resp. the right child) of node $v$. 
We build rank dictionaries for the bit arrays $b^c_v$. 

$A^c_{\mbox{{\scriptsize{\em left}}}(v)}$ and $A^c_{\mbox{{\scriptsize{\em right}}}(v)}$ are constructed by moving each element of $A^c_{v}$ to either $\mbox{{\em left}}(v)$ or $\mbox{{\em right}}(v)$ while keeping 
the order of elements in $A^c_v$. 
This is performed by taking into account the fact that each element of $A^c_v$ is a descriptor ID in $I^c_v$ 
satisfying two conditions: (i) $I^c_{\mbox{{\scriptsize{\em left}}}(v)} \cup I^c_{\mbox{{\scriptsize{\em right}}}(v)} = I^c_v$ and (ii) $I^c_{left(v)} \cap I^c_{right(v)} = \emptyset$. 
Bit $b_v^c[k]$ indicates $A^c_v[k]$ moves to whether $left(v)$ or $right(v)$.
$b_v^c[k]=0$ indicates $A^c_{v}[k]$ moves to $A^c_{left(v)}[k]$ and 
$b_v^c[k]=1$ indicates $A^c_{right(v)}[k]$ inherits $A^c_{v}[k]$. 
Bit $b_v^c[k]$ is computed by $A_v^c[k]$ as follows:
\[
  b_v^c[k] = \left\{
\begin{array}{ll}
  1 & \mbox{if $A^c_v[k] > \lfloor (a+b)/2 \rfloor$} \\
  0 & \mbox{if $A^c_v[k] \leq \lfloor (a+b)/2 \rfloor$}.
\end{array}
\right.
\]
Figure~\ref{fig:sitad} shows bit $b^{10}_{root}$, $b^{10}_{left(root)}$ and $b^{10}_{right(root)}$ computed from 
$A^{10}_{root}$, $A^{10}_{left(root)}$ and $A^{10}_{right(root)}$, respectively.  
For example, $b^{10}_{root}[7]=0$ indicates $A^{10}_{root}[7]=A^{10}_{left(root)}[5]=1$.
$A^{10}_{root}[8]=A^{10}_{right(root)}[3]=6$ is indicated by $b^{10}_{root}[8]=1$.

To perform a similarity search for query of $m$ non-zero weights $Q=(d_1:f_1,d_2:f_2,\cdots ,d_m:f_m)$, 
SITAd computes $s_{vj}$ and $t_{vj}$ at each node $v$ and checks constraint~(\ref{eq:cost}) by computing $I[s_{vj} \leq t_{tj}]$ 
and $\max E_{root}(s_{vj},t_{vj})$ on RMQ data structure $U^{c}$. 
SITAd sets $s_{vj}=P^c_v[d_j-1]+1$ and $t_{vj}=P^c_v[d_j]$ only at the root $v$.
Using $s_{vj}$ and $t_{vj}$, 
SITAd computes $s_{left(v)j}$, $t_{left(v)j}$, $s_{right(v)j}$ and $s_{right(v)j}$ by using rank operations in $O(1)$ time as follows:
\begin{eqnarray}
  &s_{left(v)j} =& rank_0(b_v^c, s_{vj} -1),~~ t_{left(v)j} = rank_0(b_v^c,t_{vj}) \nonumber \\
  &s_{right(v)j} =& rank_1(b_v^c, s_{vj}-1) + 1,~~ t_{right(v)j} = rank_1(b_v^c, t_{vj}). \nonumber
\end{eqnarray}
Note that $P^c_v$ is required at the root for maintaining $s_{vj}$ and $t_{vj}$. 
Thus, SITAd keeps $P^c_v$ only at the root. 

The memory for storing $b^c_v$ for all nodes in $T^c$ is $N^c\log{|B^c|} + o(N^c\log{|B^c|})$ bits.
Thus, SITAd needs $\sum_{c=1}^D N^c(\log{|B^c|}+\frac{N^c\log{N^c}}{2}+N^c\log{M}+2N^c+o(N^c\log{|B^c|}))$ bits of space for storing $b^c_v$ at all nodes $v$ and an RMQ data structure $U^c$ for all $c \in [1,D]$.
The memory requirement of SITAd is much less than that for storing summary descriptors $y^c_v$ using $D\log{M}\sum_{c=1}^{D}|B^c|\log{(|B^c|)}$ bits. In our experiments $D=642,297$. 
Although storing $P^c_{root}$ needs $D\sum_{c=1}^D\log{N^c}$ bits, this is not an obstacle in practice, even for large $D$.

\section{Experiments}
\subsection{Setup}
We implemented SITAd and compared its performance to the following alternative similarity search methods: one-vs-all search (OVA); an uncompressed inverted index (INV); an inverted index compressed with variable-byte codes (INV-VBYTE); an inverted index compressed with PForDelta codes (INV-PD).
All experiments were carried out on a single core of a quad-core Intel Xeon CPU E5-2680 (2.8GHz). 
OVA is a strawman baseline that computes generalized Jaccard similarity between the query and every descriptor in a database. 
INV was first proposed as a tool for cheminformatics similarity search of molecular descriptors by Kristensen et al.~\cite{Kristensen11} 
and is the current state-of-the-art approach. INV-VBYTE and INV-PD are the same as INV except that the inverted lists are compressed using variable-byte codes and PForDelta, respectively, reducing space requirements. 
We implemented these three inverted indexes in C++. For computing rank operations in SITAd we used an efficient implementation of hybrid bitvector~\cite{Kark14a} downloadable from \url{https://www.cs.helsinki.fi/group/pads/hybrid_bitvector.html}. 

Our database consisted of the 42,971,672 chemical compounds in the PubChem database~\cite{Chen09}. 
We represented each compound by a descriptor with the dimension of 642,297 constructed by the KCF-S algorithm~\cite{Kotera12}. 
We randomly sampled 1,000 compounds as queries. 

\subsection{Results}

\begin{table}[t]
\begin{center}
\caption{Performance summary showing average search time, memory in megabytes (MB), number of selected blocks per query (\#$B^c$), average number of traversed nodes (\#TN), and average number of rank computations (\#Ranks), when processing the database of 42,971,672 descriptors.}
\begin{tabular}{|l|c|c|c|c|}
\hline 
         &  \multicolumn{3}{c|}{{\bf Time (sec)}}              &                  \\
\cline{1-4}
         & $\epsilon=0.98$ & $\epsilon=0.95$ & $\epsilon=0.9 $ & {\bf Memory~(MB)}\\
\hline
INV  & \multicolumn{3}{c|}{$1.38 \pm 0.46$}  & $33,012$ \\
\hline
INV-VBYTE & \multicolumn{3}{c|}{$5.59 \pm 2.66$} &  $1,815$ \\
\hline
INV-PD  & \multicolumn{3}{c|}{$5.24 \pm 2.45$} & $1,694$ \\ 
\hline
OVA & \multicolumn{3}{c|}{$9.58 \pm 2.08$} &$8,171$ \\
\hline
SITAd & $0.23 \pm 0.23$ & $0.61 \pm 0.57$ & $1.54 \pm 1.47$  & $2,470$ \\
\hline\hline
\#$B^c$ & $2$   & $6$            & $12$             &          \\    
\cline{1-4}
\#TN &  $43,206$  & $118,368$  & $279,335$       &          \\    
\cline{1-4}
\#Ranks & $1,063,113$ & $2,914,208$ & $6,786,619$ & \\
\cline{1-4}
$|{\it I_N}|$ & $31$ & $132$ & $721$  &         \\     
\hline
\end{tabular}
\label{tab:total}
\end{center}
\end{table}

\begin{figure}[t]
\begin{minipage}{0.5\hsize}
\begin{center}
\includegraphics[width=0.98\textwidth]{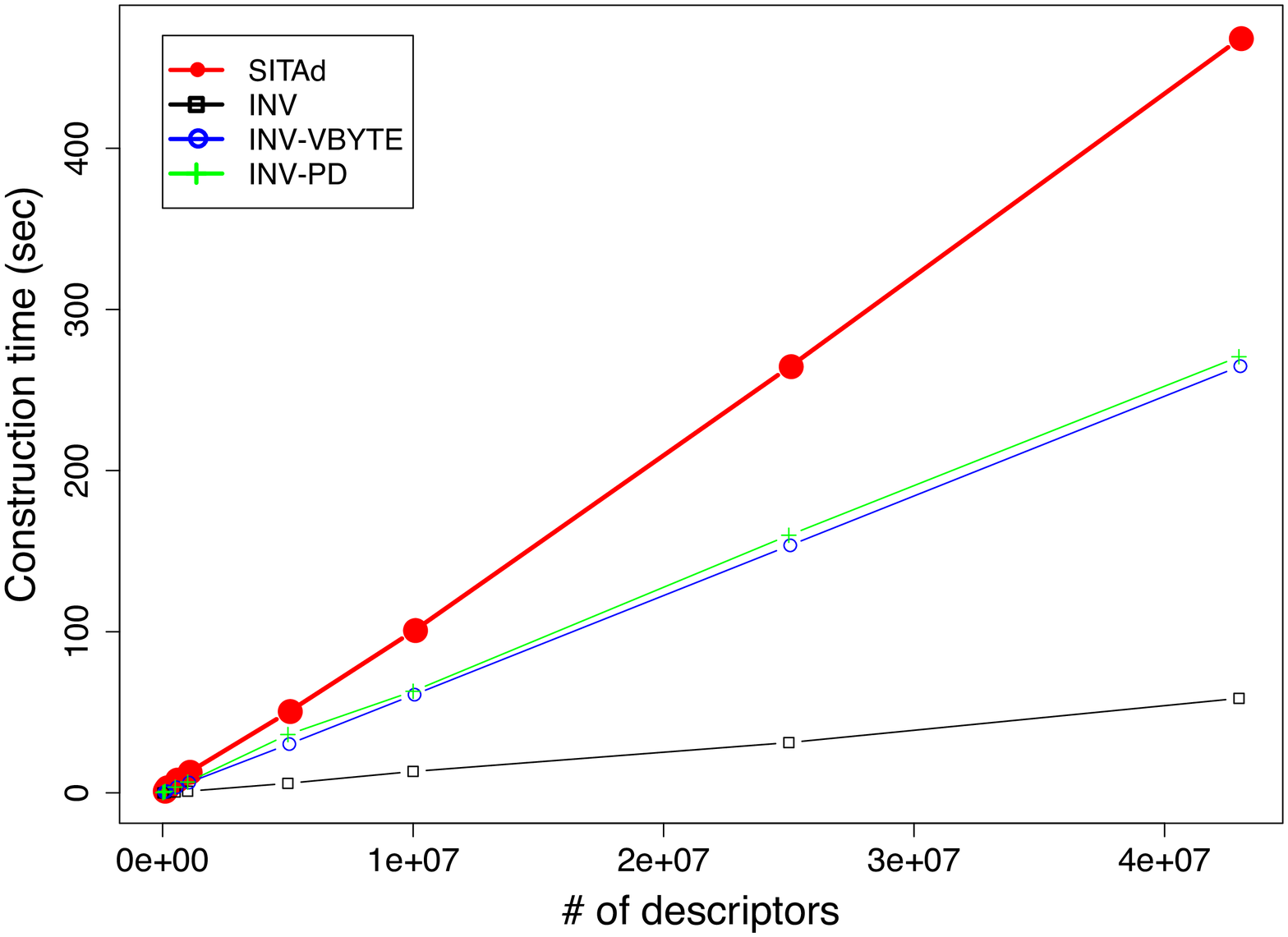}
\vspace{-0.5cm}
\end{center}
\caption{Index construction time.}
\label{fig:buildtime}
\end{minipage}
\begin{minipage}{0.5\hsize}
\begin{center}
\includegraphics[width=0.98\textwidth]{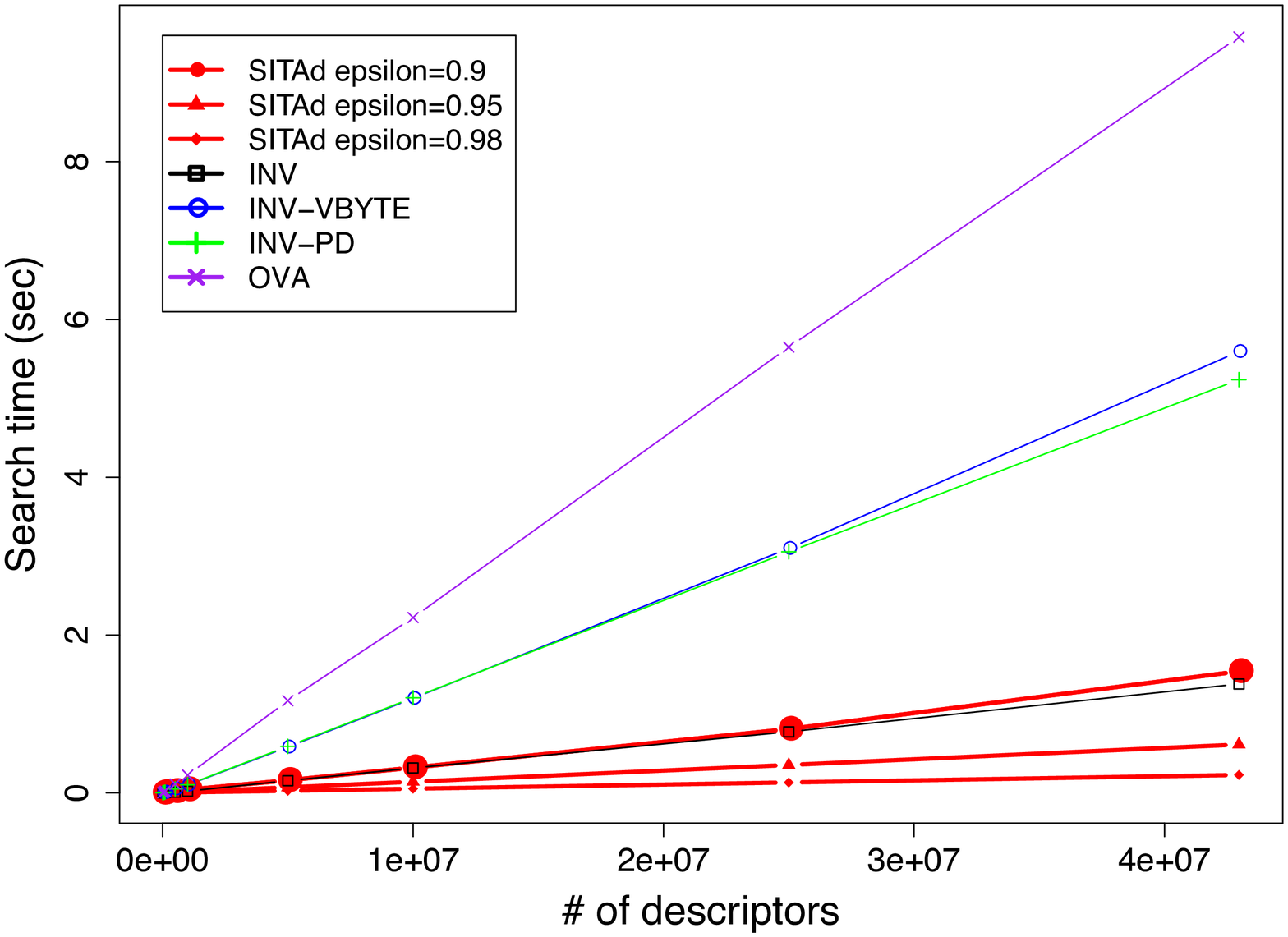}
\vspace{-0.5cm}
\end{center}
\caption{Average search time.}
\label{fig:searchtime}
\end{minipage}
\end{figure}





Figure~\ref{fig:buildtime} shows the preprocessing time taken to construct the SITAd index. 
The construction time clearly increases linearly as the number of descriptors increases, and takes only 
eight minutes for the whole database of around 42 million compounds. We should emphasize that index construction is
performed only once for a given database and that phase does not need to be repeated for each query.
Indeed, this fast construction time is an attractive and practical aspect of SITAd.

\begin{figure}
\begin{minipage}{0.5\hsize}
\begin{center}
\includegraphics[width=0.98\textwidth]{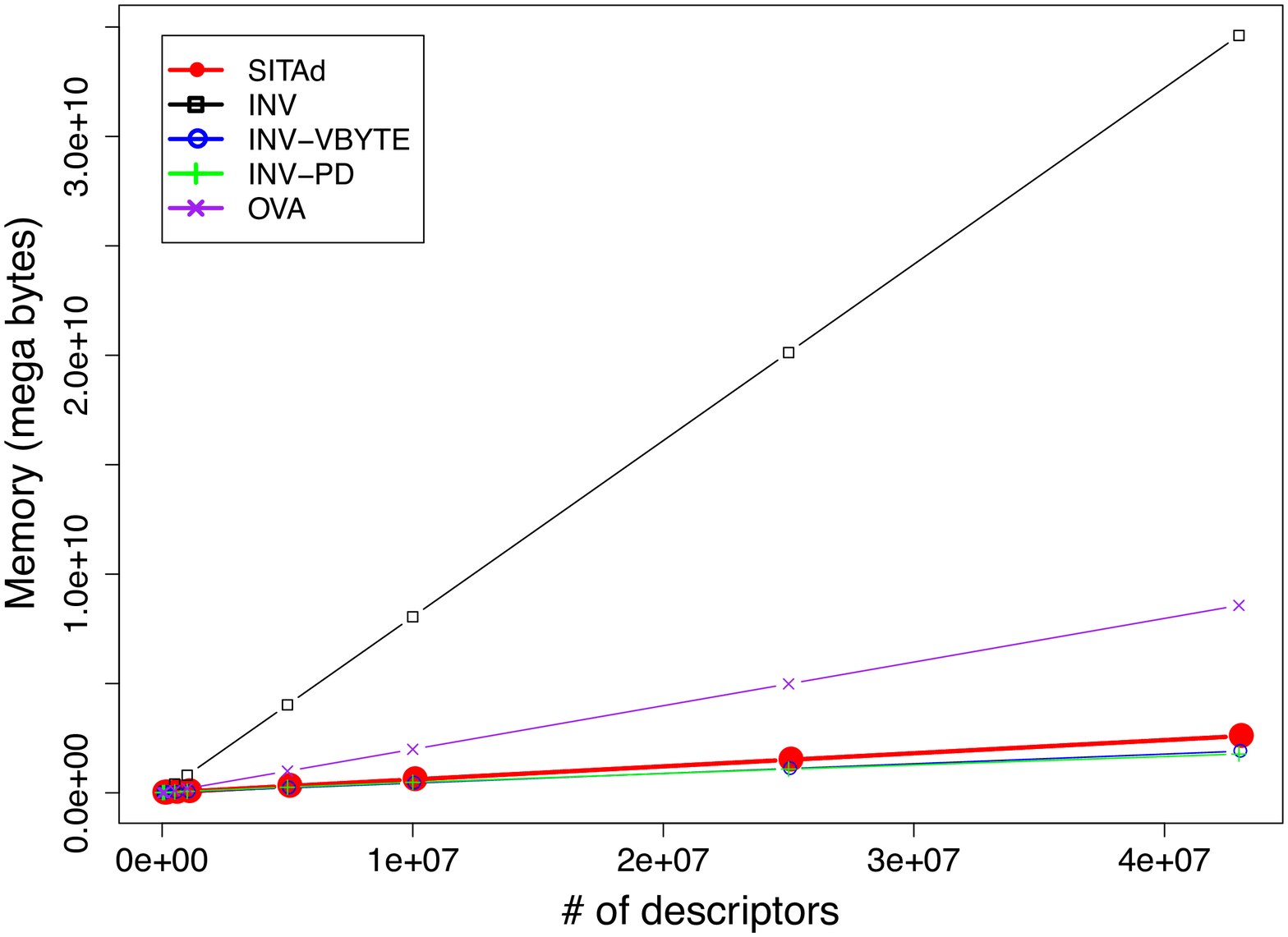}
\vspace{-0.5cm}
\end{center}
\caption{Index size.}
\label{fig:mem}
\end{minipage}
\begin{minipage}{0.5\hsize}
\begin{center}
\includegraphics[width=0.98\textwidth]{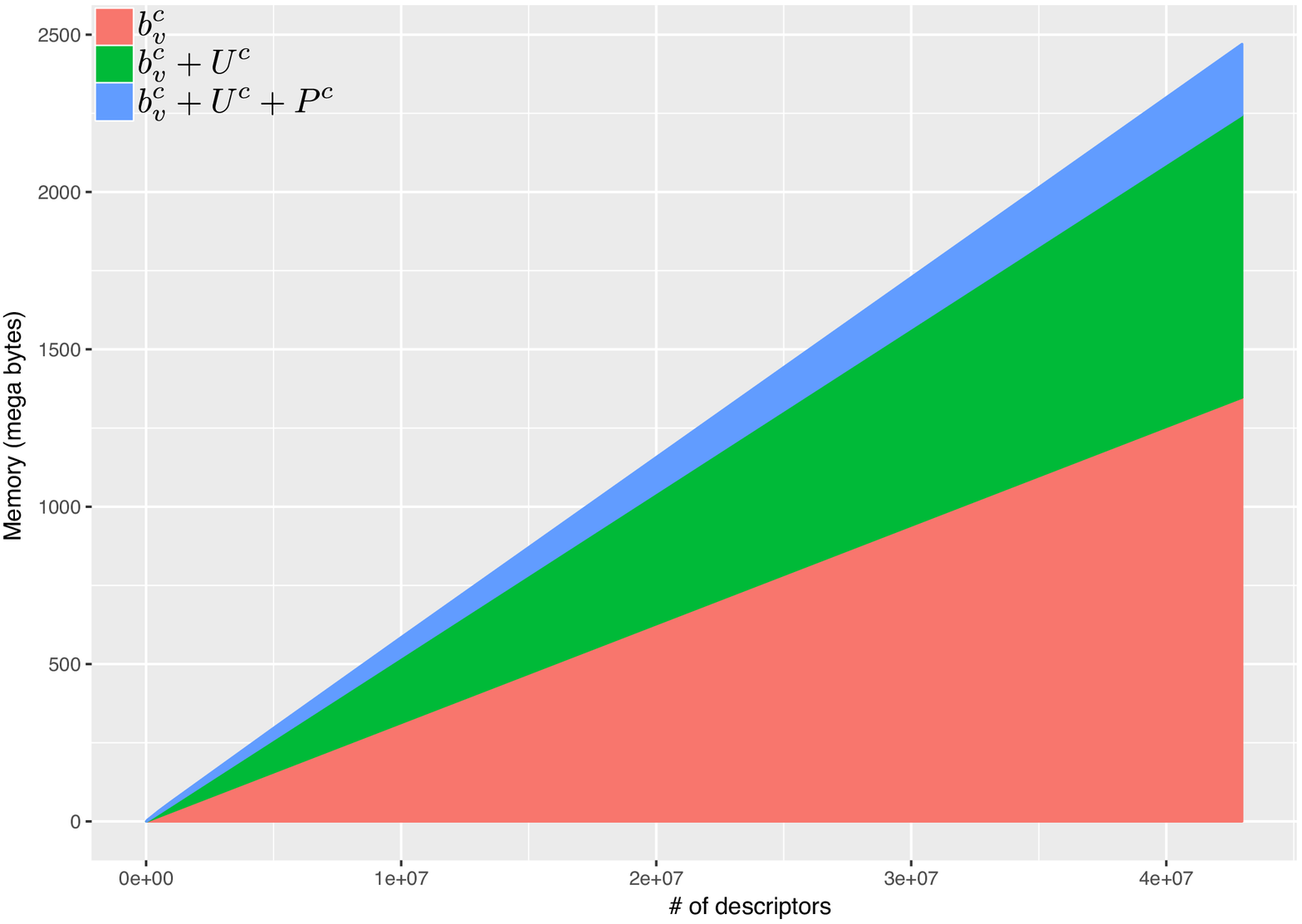}
\end{center}
\caption{Space usage of SITAd components.}
\label{fig:memsitad}
\end{minipage}
\end{figure}

Table~\ref{tab:total} shows the results of each algorithm for $\epsilon \in$ \{$0.9, 0.95$, $0.98$\}.
The reported search times are averages taken over 1,000 queries (standard deviations are also provided, and as well as small deviations), where \#$B^c$ is the number of selected blocks per query, 
$\#TN$ is the number of traversed nodes in SITAd, \#Ranks is the number of rank operations performed, and $|I_N|$ is the size of the answer set. 

Unsurprisingly OVA had the slowest search time among the tested methods, requiring 9.58 seconds per query on average and using $8$GB of main memory. 
In line with previously reported results\cite{Kristensen11}, INV provided faster querying than OVA but used more memory. The average search time of INV was faster than that of SITAd when the latter system had $\epsilon=0.9$, but became significantly slower than SITAd with large $\epsilon$ of $0.95$ and $0.98$. 
INV required $33$GB of main memory, the most of any system.
The compressed inverted indexes INV-VBYTE and INV-PT used much smaller amounts of memory --- $1.8$ GB and $1.7$ GB, respectively. 
This space saving comes at a price, however; the average search time of INV-VBYTE and INV-PT is 4-5 times slower than that of INV. 

Overall, SITAd performed well; its similarity search was fastest for $\epsilon=0.95$ and $0.98$ and its memory usage was low.
In fact, SITAd with $\epsilon=0.98$ was $20$ times faster than INV-VBYTE and INV-PD with almost the same memory consumption. 
It took only $0.23$ and $0.61$ seconds for $\epsilon= 0.98$ and $0.95$, respectively, and it used small memory of only $2$ GB, which fits into the memory of an ordinary laptop computer. 
Its performance of SITAd was validated by the values of \#$B^c$, \#TN and \#Ranks. 
The larger the threshold $\epsilon$ was, the smaller those values were, which demonstrates efficiency in the methods for pruning the search space in SITAd. 

Figure~\ref{fig:searchtime} shows that for each method, the average search time per query increases linearly as the number of descriptors in the database increases. Figure~\ref{fig:mem} shows a similar linear trend for index size. Figure~\ref{fig:memsitad} illustrates that for SITAd, rank dictionaries of bit strings $b^c_v$ and RMQ data structure $U^c$ are the most space consuming components of the index.

\section{Conclusion}
We have presented a time- and space-efficient index for for solving similarity search problems that we call the {\em succinct intervals-splitting tree algorithm for 
molecular descriptors (SITAd)}. It is a novel, fast, and memory-efficient index for generalized-Jaccard similarity 
search over databases of molecular compounds. The index performs very well in practice providing speeds
at least as fast as previous state-of-the-art methods, while using an order of magnitude less memory.
In future work we aim to develop and deploy a software system using SITAd, which will be of immediate
benefit to practitioners.



\bibliographystyle{abbrv}
\bibliography{biblio} 

\end{document}